\documentclass[preprint,number,twocolumn,3p]{elsarticle}
\usepackage{url}         
\usepackage{hyperref}   
\RequirePackage{xspace}  
\usepackage{relsize}     
\include{dlsymbols}      
\usepackage{graphicx}   
\usepackage{lineno}
\newif\ifblackandwhite
\blackandwhitefalse  

\begin{document}

\title{Measured Radiation and Background Levels During Transmission of Megawatt Electron Beams Through Millimeter Apertures}

\author[ASU]{R. Alarcon}
\author[ASU]{S. Balascuta}
\author[FEL]{S.V. Benson}
\author[LNS]{W. Bertozzi}
\author[FEL]{J.R. Boyce}
\author[LNS]{R. Cowan\corref{cor1}\fnref{fn1}}
\ead{rcowan@mit.edu}
\author[FEL]{D. Douglas}
\author[FEL]{P. Evtushenko}
\author[LNS]{P. Fisher}
\author[LNS]{E. Ihloff}
\author[HMP]{N. Kalantarians}
\author[LNS]{A. Kelleher}
\author[CWM]{W.J. Kossler}
\author[FEL]{R. Legg}
\author[UNH]{E. Long}
\author[LNS]{R.G. Milner}
\author[FEL]{G.R. Neil}
\author[LNS]{L. Ou}
\author[LNS]{B. Schmookler}
\author[FEL]{C. Tennant}
\author[LNS]{C. Tschal\"ar}
\author[FEL]{G.P. Williams}
\author[FEL]{S. Zhang}

\cortext[cor1]{Corresponding author. Tel.: +1 650 926 4560.}
\fntext[fn1]{Postal address: Mail Stop 94, Stanford National Accelerator Laboratory, 2575 Sand Hill Road, Menlo Park, CA 94025 USA}

\address[ASU]{Department of Physics, Arizona State University, Glendale, AZ 85306 USA}
\address[FEL]{Free Electron Laser Group, Thomas Jefferson National Accelerator Facility, Newport News, VA 23606 USA}
\address[LNS]{Laboratory for Nuclear Science, Massachussetts Institute of Technology, Cambridge, MA 02139 USA}
\address[HMP]{Department of Physics, Hampton University, Hampton, VA 23668 USA}
\address[CWM]{Department of Physics, College of William and Mary, Williamsburg, VA 23185 USA}
\address[UNH]{Department of Physics, University of New Hampshire, Durham, NH 03824 USA}

\begin{abstract}
We report measurements of photon and neutron radiation
levels observed while transmitting a 0.43 MW electron beam 
through millimeter-sized apertures and during beam-off, but
accelerating gradient RF-on, operation.  
These measurements were conducted at the Free-Electron Laser (FEL)
facility of the Jefferson National Accelerator Laboratory (JLab)
using a 100~\mev electron beam from an energy-recovery linear accelerator.
The beam 
was directed successively through 6~mm, 4~mm, and 
2~mm diameter apertures of length 127~mm in aluminum 
at a maximum current of 4.3~mA (430 kW beam power).
This study was conducted to characterize radiation levels 
for experiments that need to operate in this environment, 
such as the proposed DarkLight Experiment. We find that
sustained transmission of a 430 kW continuous-wave (CW) beam through a 2~mm
aperture is feasible with manageable beam-related
backgrounds.  We also find that 
during beam-off, RF-on operation, 
multipactoring inside the niobium cavities of 
the accelerator cryomodules is the primary source of 
ambient radiation when the machine is tuned for 130~\mev operation.

\end{abstract}


\maketitle

%
%

\section{Introduction}

\begin{figure}[htb]
\begin{center}
\ifblackandwhite
\includegraphics[width=0.5\textwidth]{FIG1_Aprime_reach_IF_report_black_and_white.pdf}\else
\includegraphics[width=0.5\textwidth]{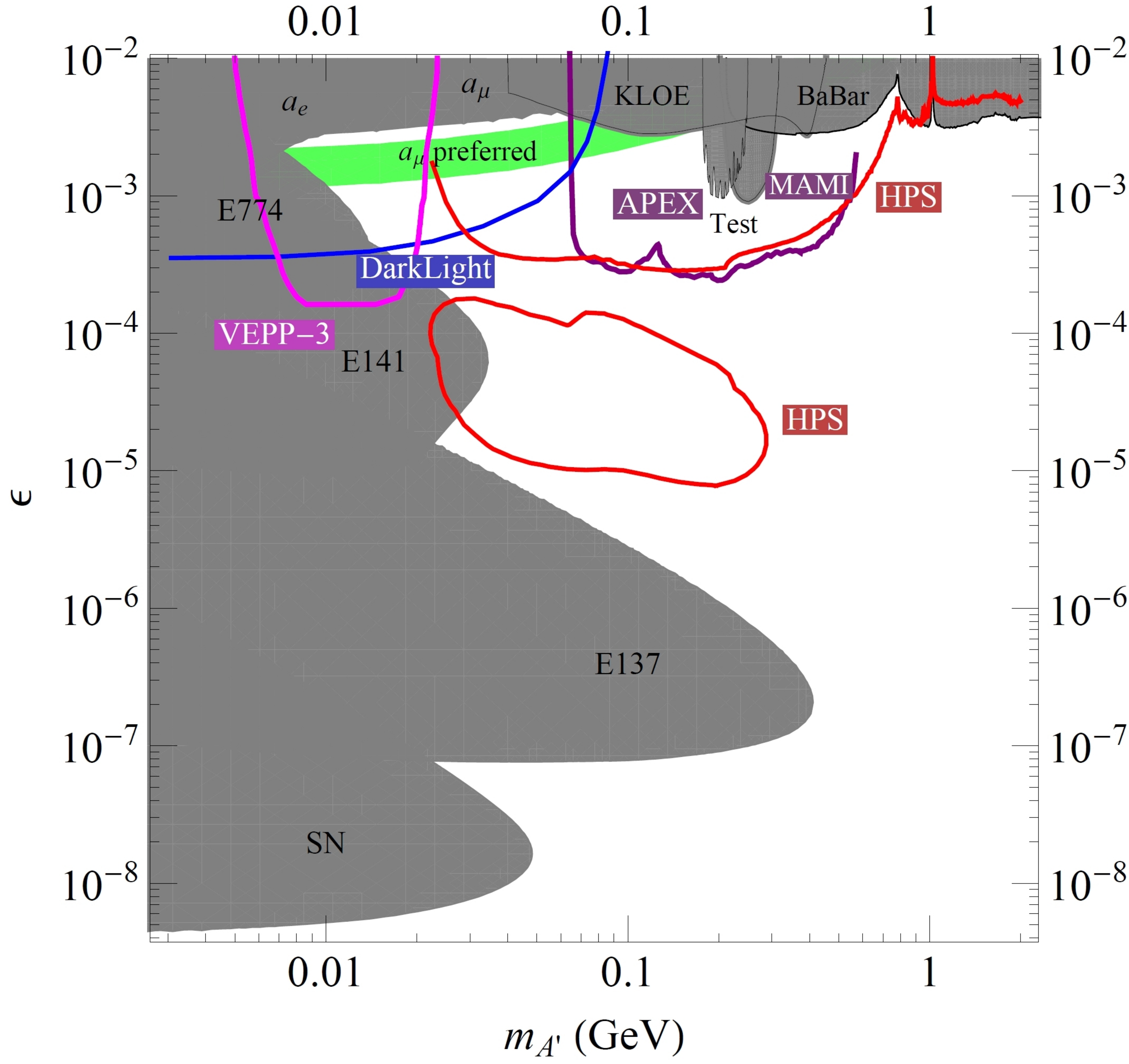}\fi
\caption{\label{ParamSpace}%
Heavy photon parameter space showing constraints from existing experiments (filled regions)
and expected reach of proposed experiments (open regions), as a function of \Apr mass and coupling 
strength $\epsilon^2=\alpha^\prime/\alpha$ to the Standard Model photon. 
APEX, DarkLight, and HPS are 
proposed experiments at JLab. From Ref.~\cite{Hewett:2012ns}.}
\end{center}
\end{figure}

The high-quality electron beam capabilities of JLab's accelerators at CEBAF and 
the Free-Electron Laser facility~\cite{Neil:2005jy}  have been incorporated into dark matter detection proposals based on predictions of Freytsis {\it et al.}~\cite{Freytsis:2009bh}. Taken together, the experiments will search for a scalar boson \Apr in the mass range $\approx 10$~\mev to 1.0~\gev.  Three experiments 
(APEX~\cite{Abrahamyan:2011gv}, 
HPS~\cite{Adrian:2013zy,HPSwebsite},
and DarkLight~\cite{FisherPAC39}) 
are either underway or are preparing to explore complementary regions of parameter space as indicated in Fig.~\ref{ParamSpace}. 

The DarkLight experiment plans to study \en-\proton scattering
in a windowless, internal gas target at the JLab FEL.
Beam-related and ambient background radiation levels at the FEL
need to be characterized.  
%
%
Here we report measurements performed 
in spring and summer 2012 including a
high-power beam transmission test through millimeter-diameter apertures.
This demonstrated that transmission of a 
half-megawatt electron beam through a 2~mm diameter, 127~mm long
aperture in an aluminum block (simulating
operation of the DarkLight gas target)
was feasible without excessive energy loss
from resistive wall heating (wakefield effects)
or from beam halo interception in the block.  It also allowed
photon and neutron radiation levels to be studied.  Companion 
papers~\cite{Tschalaer:2013ab,Alarcon:2013yxa} address the beam loss and heating effects; here,
we address the radiation measurements.

\section {Experimental set-up}

\begin{figure}[tbh]
\hbox to\hsize{\hss
\ifblackandwhite
\includegraphics[width=0.5\textwidth]{FIG2_NAI-PMT_with_extra_lead_17Jul2012_DSCN0075_black_and_white.pdf}\hss}\else
\includegraphics[width=0.5\textwidth]{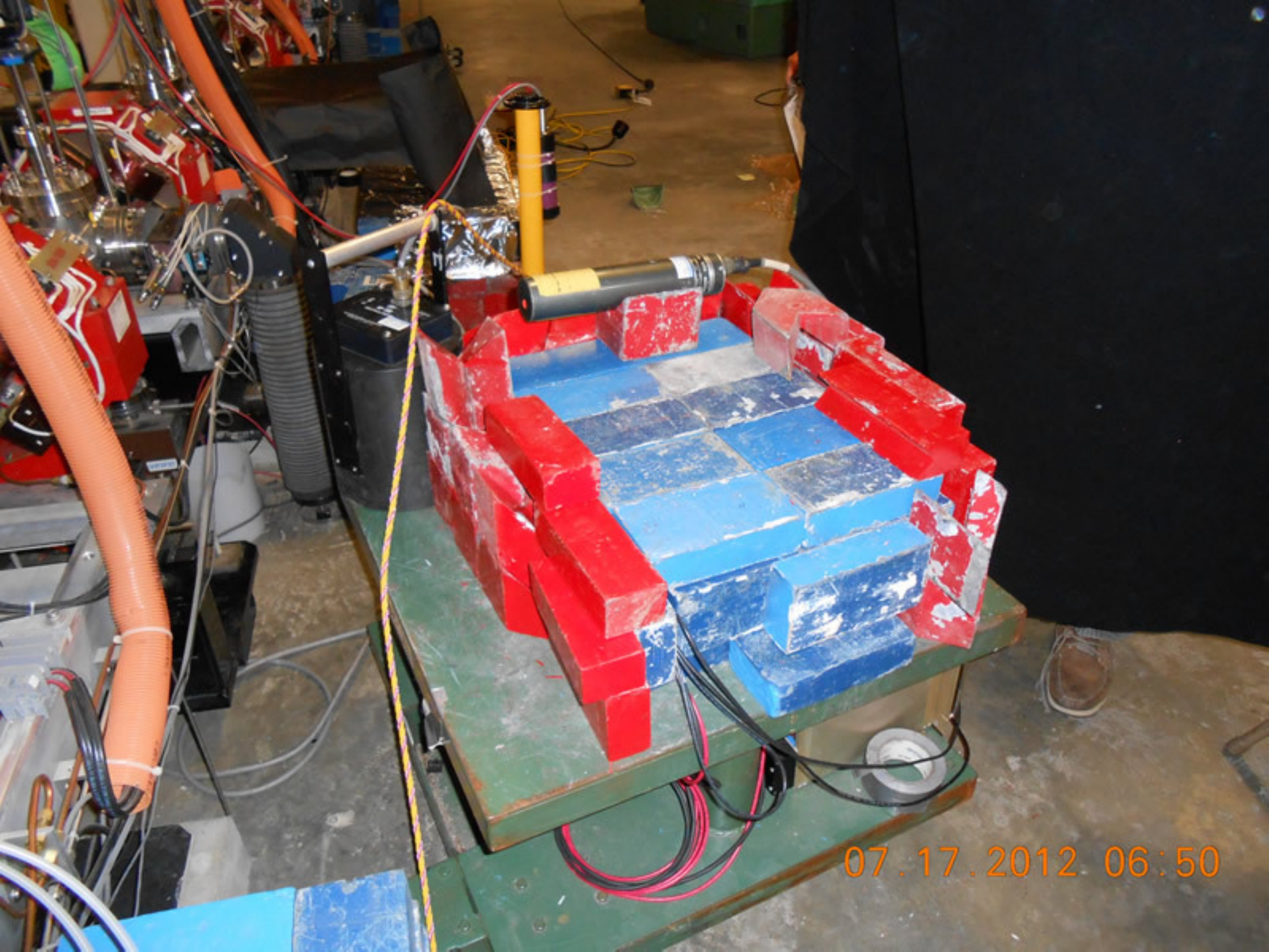}\hss}\fi
\caption{\label{fig:extra_lead}%
The three detector systems on the movable cart.  The two NaI/PMT detectors
are housed inside the shielding enclosure constructed of 
lead bricks.  Sufficient lead was used to reduce the dead time during
430 kW running to manageable levels.
The test fixture containing the aperture block is in the upper left,
but is largely obscured by other beam line components.  
The neutron monitor RM212\_p1, is the cylinder on the left-most corner of the
cart, and the photon monitor RM212\_p2 is seen on top of the lead stack.}
\end{figure}

Measurements were carried out in the shielded vault of the Jefferson Laboratory
FEL.  The FEL consists of an injection system including a drive laser, an energy recovery linear accelerator (ERL) with
superconducting RF cavities, beam lines, and recirculation arcs.  The ERL 
produces an electron
beam of up to 130~\mev energy and up to 1~MW power.  The electron beam can drive
either of two beam lines.  One is configured to produce infrared (IR) light,
the other to produce ultraviolet (UV) light.  After traversing either beam line, 
the beam is returned to the ERL for energy extraction. 

For the transmission test, a 6-port vacuum chamber test fixture 
was constructed containing a movable aluminum target block with three apertures
of 6~mm, 4~mm, and 2~mm diameter, each 127~mm in length.   The 2~mm
aperture approximates
the geometry of the entrance and exit ports of the proposed DarkLight
windowless gas target, where the narrow aperture is used to restrict gas flow 
from the target into the beam line vacuum.  
Diagnostics were fitted on the aperture block, 
including a temperature sensor, a YAG crystal with a $10^5$ dynamic range
for measuring the beam halo, and an optical transition radiation (OTR)
silicon sensor for measuring the beam profile.  
Additional beam line diagnostics and viewers were used as well.
For the transmission test, the fixture containing the aluminum aperture block 
was installed in the FEL IR beam line, 
along with additional quadrupoles and other beam line
components.  Ambient radiation measurements (those without the test fixture installed)
were made using the IR and UV beam lines.

\begin{figure*}[htb]
\begin{center}
\ifblackandwhite
\includegraphics[width=\textwidth]{FIG3_FEL-RAS-layout_v2_black_and_white.pdf}\else
\includegraphics[width=\textwidth]{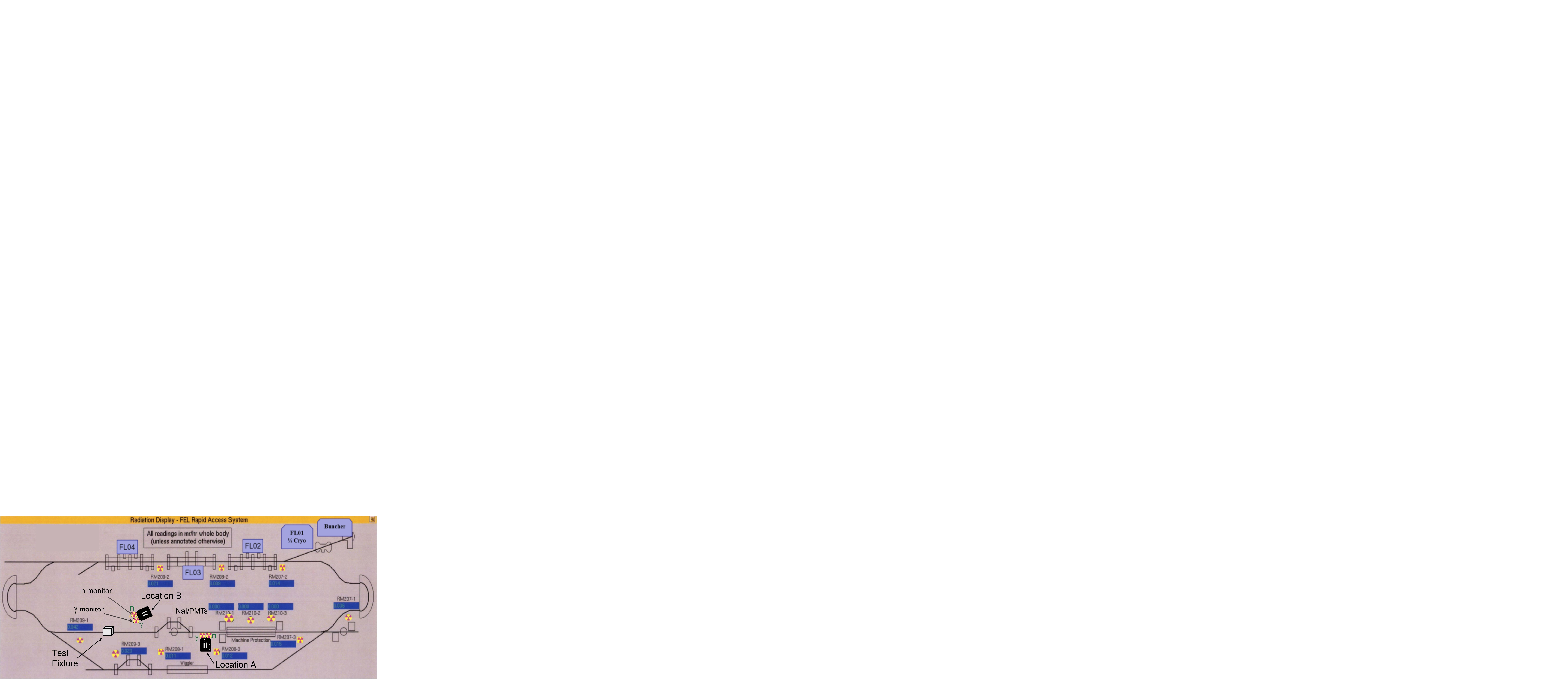}\fi
\caption{\label{FEL-RAS-layout}
FEL Rapid Access System lay-out on the vault floor plan. Radiation symbols indicate the location of a Radiation Controls department (RadCon) radiation monitor. The two NaI/PMT detectors inside their lead shielding, and the two RadCon monitors RM-212\_p1 and RM-212\_p2 were first located at Location~A, near the middle of the vault, for ambient radiation measurements conducted during UV lasing runs. When the test fixture with the aperture block was installed, the four detectors were relocated to Location~B.}
\end{center}
\end{figure*}

Three radiation monitoring systems were installed in the FEL vault for this study: a pair of NaI/PMT detectors (with 2-inch and 3-inch diameter NaI crystals, respectively), each inside a lead-shielded enclosure constructed with standard 2-inch x 4-inch x 8-inch lead bricks, an unshielded neutron monitor (Canberra NP100B counter, designated RM212\_p1) and an unshielded gamma monitor, a xenon-filled ion chamber (Canberra IP100, designated RM212\_p2).  The NaI/PMT detectors were calibrated and adjusted to count photons from 300~\kev to 15~\mev.
All three systems were located on a movable cart
(see Fig.~\ref{fig:extra_lead}).  The NaI/PMT detectors 
were read out by spectroscopy amplifiers at the cart, with high-level signals sent to MCAs in laboratory space outside the vault.
The
gamma and neutron monitors were read out by the FEL radiation protection
monitoring system.
Table~\ref{tab:detectors_monitors} lists the detectors used in this study.
See Fig.~\ref{FEL-RAS-layout} for their locations in the vault and relative to
the test fixture, IR and UV beam lines, and linac cryomodules.

\begin{figure}[htb]
\begin{center}
\ifblackandwhite
\includegraphics[width=0.5\textwidth]{FIG4_NaI_PMT_graphic_revised_v2_cropped_black_and_white.pdf}\else
\includegraphics[width=0.5\textwidth]{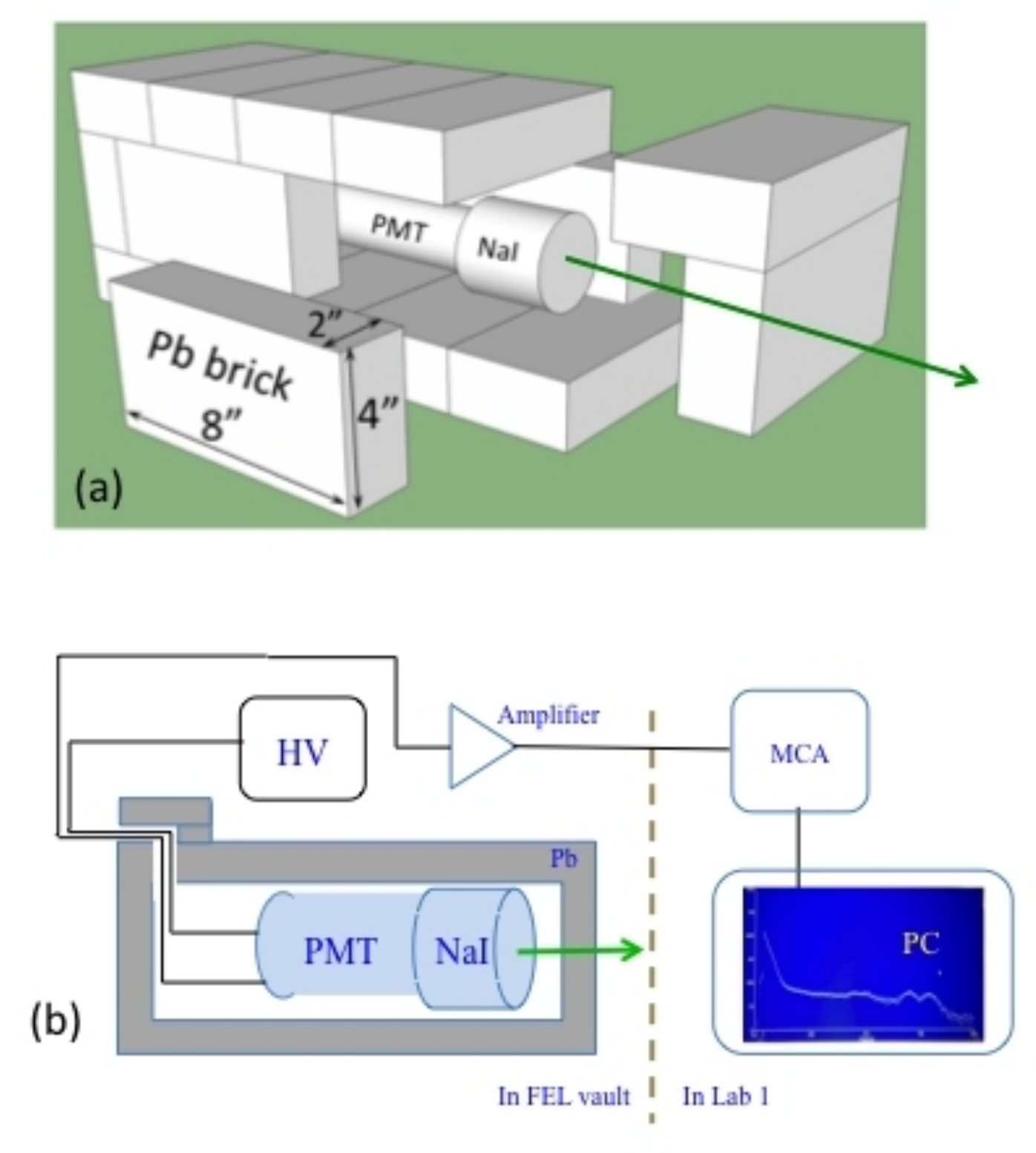}\fi
\caption{\label{Figure1A}Ambient photon radiation was recorded by a NaI/PMT detector inside a lead box. Upper plot (a) shows an exploded view of the basic setup for a 2~inch thick lead enclosure constructed with standard 2-inch x 4-inch x 8-inch lead bricks. Lower plot (b) shows the data acquisition circuit for the NaI/PMT system. Two independent NaI/PMT systems were used: one 2~inches dia. x 2~inches in length, the second 3~inches dia. x 3~inches in length. Both systems were calibrated using \textsuperscript{60}Co ( $\gamma$ = 1.17 MeV and 1.33 MeV) and \textsuperscript{137}Cs ($\gamma$ = 0.6617 MeV) sources.}
\end{center}
\end{figure}

Ambient background data were taken while the
cart was positioned near the mid-point
of the IR beam line (Location~A in Fig.~\ref{FEL-RAS-layout}).
Radiation data with the test fixture installed were
taken with the cart at a location further
upstream on the IR line and approximately 1.9~m downstream of the test fixture,
at $24^{\circ}$ to the side (Location~B).
Fig.~\ref{Figure1A} shows the shielding configuration for the NaI/PMT detectors. 

\begin{table}
\hbox to\hsize{\hss\smaller
\begin{tabular}{|c|c|l|}
\hline
\hline
Monitor & Type & Location during transmission test\\
\hline
\hline
RM212\_p1 & neutron & 1.9~m downstream of test fixture \\
RM212\_p2 & photon &  1.9~m downstream of test fixture \\
2 inch NaI & --- & 1.9~m downstream of test fixture \\
3 inch NaI & --- & 1.9~m downstream of test fixture \\
\hline
\hline
\end{tabular}\hss}
\caption{\label{tab:detectors_monitors}
Radiation detectors used in this study. Their locations in the FEL vault
are shown in Fig.~\protect\ref{FEL-RAS-layout}.}
\end{table}

\begin{table}
\hbox to\hsize{\hss\smaller
\begin{tabular}{|c|c|r|c|r|c|}
\hline
\hline
Run & Aperture & Duration & Beam  & Charge & Average  \\
\#   & diameter &   &   power     & delivered             & current \\
\hline
\hline
1 &  6 mm & 22 min &   0.384 MW & 5.1 C & 3.84 mA\\
2 &  4 mm & 30 min &   0.393 MW & 7.1 C & 3.93 mA\\
3 &  2 mm & 124 min &  0.425 MW & 31.6 C & 4.25 mA \\
4 &  2 mm & 413 min &  0.422 MW & 121 C & 4.22 mA\\
\hline
\hline
\end{tabular}\hss}
\caption{\label{tab:four_runs}%
Running conditions for the four reference runs during the beam transmission study.}
\end{table}

\section{Radiation level measurements}

Radiation measurements were made for various machine configurations:
beam-on, with the test fixture containing the aperture block in place on the beam line; 
beam-on, without the test fixture and aperture block; and
beam-off, but with accelerating gradient RF applied to the cryomodules.

\subsection{Ambient measurements without the aperture block}

\begin{figure}[htb]
\ifblackandwhite
\includegraphics[width=0.5\textwidth]{FIG5_Figbk03_crop_black_and_white.pdf}\else
\includegraphics[width=0.5\textwidth]{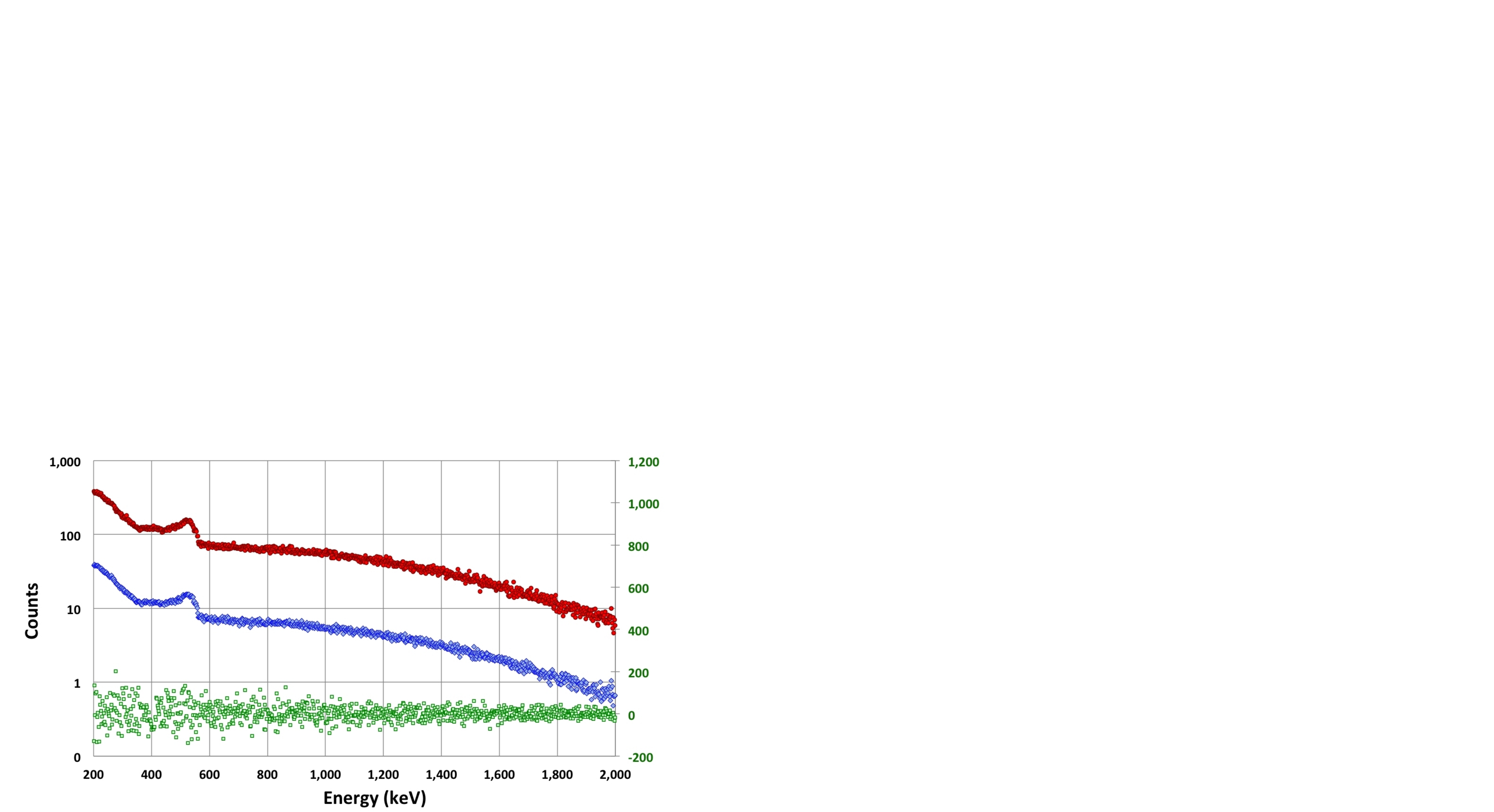}\fi
\caption{\label{TwoSpectra}%
NaI/PMT spectra taken with beam-on (middle band diamonds, vertical scale on left) 
and beam-off, RF-on (upper band circles, vertical scale on left).
Beam-off data points are plotted times 10 for clarity.
Beam-on data were taken with the FEL 
operating at 75~MHz CW, a well-tuned \en beam, and lasing in the UV.
The beam-off, but RF-on data were taken immediately
after the electron beam was turned off. 
Their difference (lower band squares, vertical scale on right) is the contribution to the total 
due to the electron beam itself.   
The other main source of ambient vault radiation is the cryomodules.}
\end{figure}

To study ambient backgrounds in the vault (without the test fixture installed), a
set of measurements was made 
during beam-on running, with electrons in the UV line, and again during
beam-off, but accelerating-gradient-on, running.  
NaI/PMT spectra obtained under these running conditions
are shown in Fig.~\ref{TwoSpectra}. The beam-on spectrum is shown in
blue, and the beam-off, RF-on spectrum in red.   
The small
difference between them (shown in green)
demonstrates 
that the ambient radiation in the FEL vault for this particular machine 
tune is not associated with the presence of electrons in the machine. 
Rather,
when the electron beam is properly tuned for maximum energy recovery
and 130~\mev beam energy, as was the case here, the 
source of the ambient radiation is multipactoring in the RF 
cryomodules~\cite{Myneni:2008zz}.  Multipactoring occurs when field 
emission pulls electrons from the niobium surface of a cryomodule
and these electrons are accelerated, then impact the cavity wall
and release more electrons and potentially photons and neutrons as well,
depending on the gradient.  The process can then repeat.

%
%

\begin{figure}[htb]
\ifblackandwhite
\includegraphics[width=0.5\textwidth]{FIG6_Whole_body_dose_rates_copy_crop_black_and_white.pdf}\else
\includegraphics[width=0.5\textwidth]{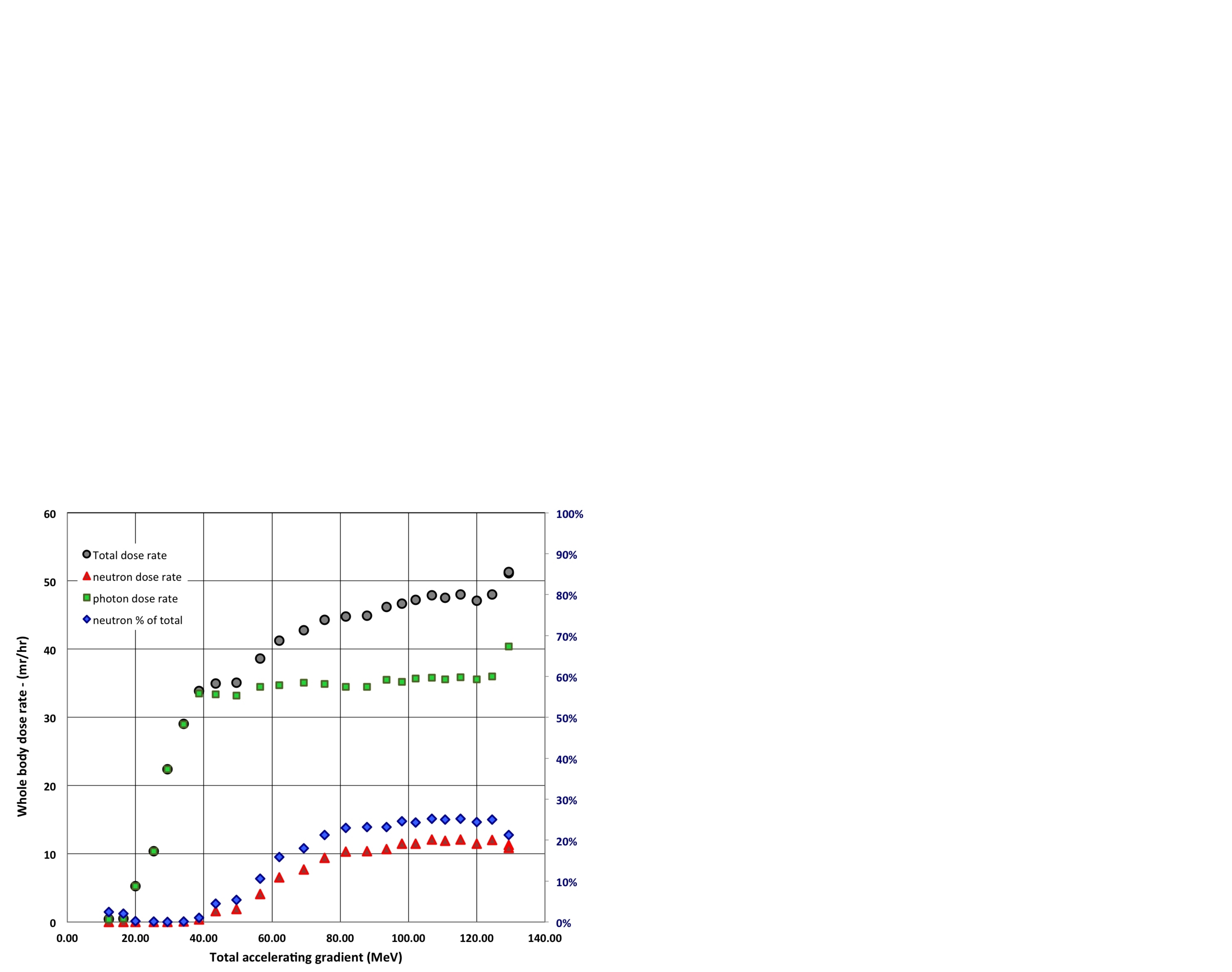}\fi
\caption{ \label{DoseRates}%
Photon (squares), neutron (triangles), and total (circles) ambient dose 
rates as a function of accelerating gradient
when no beam is present (left vertical scale).  
The neutron dose rate as percentage of the total dose
rate (diamonds, right vertical scale) is also shown.
These measurements were taken at Location~A in 
Fig.~\protect\ref{FEL-RAS-layout}.}
\end{figure}

A series of radiation measurements
were made as a function of accelerating gradient.
The RM212\_p1 (neutron) and RM212\_p2 (photon) counters
were used to measure dose rates, which are shown in 
Fig.~\ref{DoseRates} as a function of RF accelerating
gradient. 
Starting with the accelerator tuned for stable operation at 130~\mev, and with no electrons 
in the machine, 
RF cavities were turned off sequentially, working back towards the injector. 
After each section was turned off, radiation levels were recorded. 
Neutrons are seen to be 
responsible for about one quarter of the dose rates when the RF power 
is set up for accelerated energy greater than 100~\mev.  These measurements were
made with the RM212 monitors at Location~A.

We note that this setup---tuning the linac for maximum energy---tends toward the worst-case
scenario in terms of production of photon and neutron backgrounds by
multipactoring.  However, the amount of radiation can be managed by balancing
the machine tune to provide 
the required beam energy, stable energy recovery, and satisfy
any other considerations that may be present.  Also, over time, RF cavities
become more susceptible to field emission.  Replacement of cavities exhibiting
more pronounced field emission can reduce the radiation levels.

\subsection{Measurements with the aperture block}

\begin{figure*}[htb]
\begin{center}
\ifblackandwhite
\includegraphics[width=\textwidth]{FIG7_DL_Z3F_rad_data_Run_027_annotations_removed_v11_black_and_white.pdf}\else
\includegraphics[width=\textwidth]{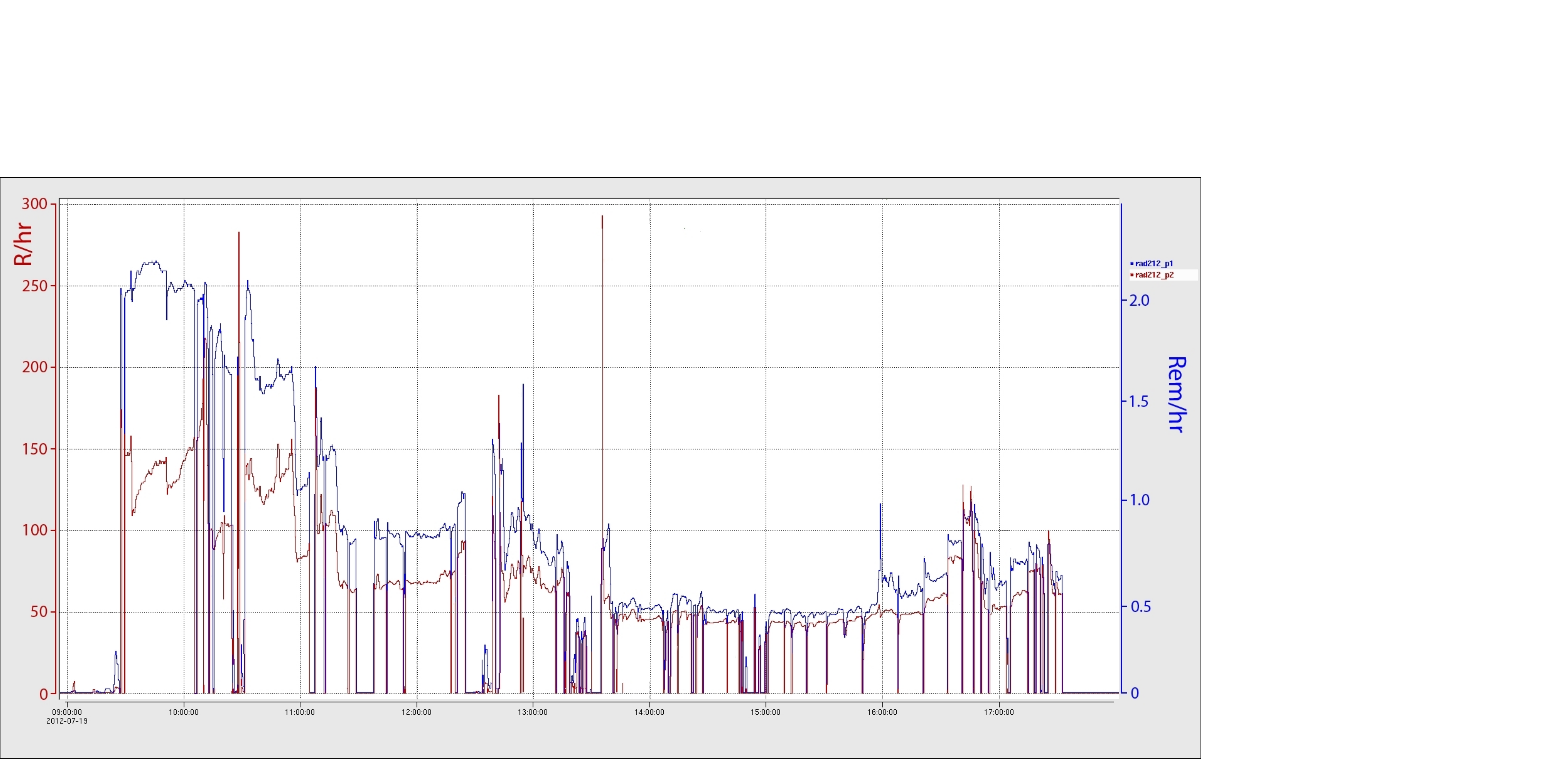}\fi
     \caption{\label{fig:rad_data}%
Photon (darker trace, left axis) and neutron (lighter trace, right axis)
radiation levels during the 7~hour Run~4.
Several short-duration trips of the machine can be seen.  Radiation levels decreased
during the run as machine tune was improved.}
\end{center}
\end{figure*}

Several runs were carried out with the test fixture and its aperture block
installed on the IR beam line.  Table~\ref{tab:four_runs} lists four of
these runs.  Starting with the 6~mm aperture and progressing
to the 4~mm and finally to the 2~mm aperture, the tests were conducted
with an initially low charge/bunch and repetition rate.
These were then increased while maintaining machine tune for 
minimum beam loss.  Radiation background data were recorded for each
test, along with the temperature change in the block and machine parameters.
The tests varied in length from 15~minutes
to 2~hours, with 
a final long run (Run~4) using the 2~mm aperture 
lasting 7~hours and delivering 121~C through the aperture.
Fig~\ref{fig:rad_data} shows
the recorded radiation levels for the photon and neutron monitors listed in
Table~\ref{tab:detectors_monitors}.

Stable operation with minimal beam loss was achieved at each aperture 
step.  The beam parameters at each step
were
100~MeV electron energy, 
53--60~pC/bunch, 
4.0--4.3 mA beam current,
75~MHz bunch frequency, 
400--430 kW beam power, and
70 $\mu$m beam spot size at the aperture block (20~cm beta function at r.m.s. emittance of 
$2.5\times 10^{-8}$).
The beam profile and halo were measured during low duty cycle, pulsed-beam
runs.

The temperature increase of the 1~kg aluminum block during
beam operation, along with the cooling rate
during beam-off time, allow calculation of the average power
deposited in the block.  When combined with the total beam power,
the fractional beam loss can be determined.  See Ref.~\cite{Tschalaer:2013ab} for a more
detailed discussion.  
Table~\ref{tab:beam_losses} shows beam losses ranging from 1.3~ppm to 6.8~ppm.

\begin{table*}
\hbox to\hsize{\hss\smaller
\begin{tabular}{|c|c|c|c|c|c|}
\hline
\hline
Run & Aperture & Run      & Average   & Neutron dose  & Photon dose \\
\#  & diameter & duration & beam loss & rate at Loc.~B & rate at Loc.~B \\
\hline
\hline
1 & 6 mm & 22 min & 1.3 ppm & 0.24 rem/hr  & 13 R/hr \\
2 & 4 mm & 30 min & 2.1 ppm & 0.43 rem/hr  & 19 R/hr \\
3 & 2 mm & 124 min & 6.8 ppm & 1.32 rem/hr  & 75 R/hr \\
4 & 2 mm & 413 min & 2.8 ppm & 0.58 rem/hr  & 60 R/hr \\
\hline
\hline
\end{tabular}\hss}
\caption{\label{tab:beam_losses}%
Average beam loss and 
radiation backgrounds (whole-body dose rate) 
observed for each aperture, averaged for each of the four reference runs.
Photon and neutron backgrounds are at Loc.~B, 1.9~m downstream of the test fixture
and 1~m to the side of the beam line.}
\end{table*}


Table~\ref{tab:beam_losses} also lists values of the photon and neutron
backgrounds.  As operational experience was gained, 
backgrounds decreased during each run.
In Run~4, the photon dose rate downstream of the
test fixture started near 150~R/hr, decreased to around 40~R/hr as the tune
was improved, then remained stable to the end of the run.
Neutron production based on beam loss calculations with the 2~mm aperture
ranged from 0.6~rem/hr to 1.3~rem/hr.

\begin{figure}[htb]
\begin{center}
\ifblackandwhite
\includegraphics[width=0.5\textwidth]{FIG8_FLUKA-Run-4_crop_black_and_white.pdf}\else
\includegraphics[width=0.5\textwidth]{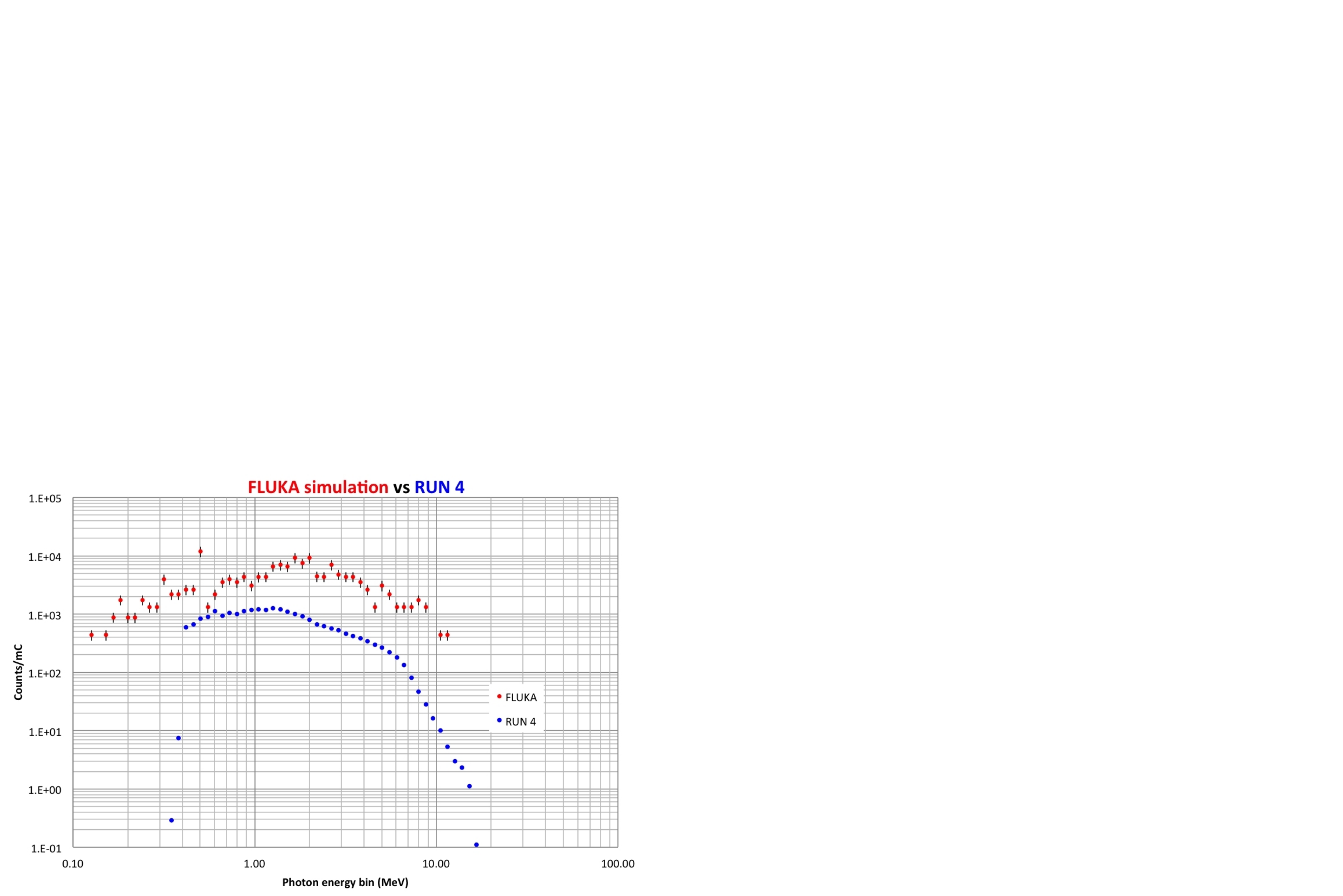}\fi
\caption{\label{fig:fluka_comparison}%
Comparison of NaI/PMT Run~4 spectrum (points without error bars) with a
FLUKA simulation (points with error bars).  The NaI/PMT data were recorded
by the 2-inch dia. NaI/PMT detector.
The FLUKA spectrum is from a simulation of
the aperture block and beam line components and models the photon
spectrum incident on the NaI/PMT detector inside the 4~inch lead shield.
Both spectra are in units of photons/mC of beam electrons incident on 
the aperture block and are normalized to the surface 
area of the NaI crystal.  The
FLUKA simulation does not include the effects of the NaI/PMT detector
response, which had a 40\% average dead time during Run~4, nor the neutron
component.}
\end{center}
\end{figure}

\section{Background Modeling}
A simulation of the transmission test using the 2~mm aperture was 
made using FLUKA\cite{Battistoni:2007zzb,Ferrari:2005zk} to compare
with the measured NaI/PMT data.  The simulation, which modeled $2.5\times 
10^{6}$ incident \en, 
included the aluminum aperture block, an approximation of the beam line
and its components to a distance of 3~m downstream of the aperture
block, and the lead enclosure around the detectors.  The FLUKA photon
spectrum inside the shielded enclosure is 
shown in Fig.~\ref{fig:fluka_comparison}
along with the NaI/PMT data.  A simulation of the NaI/PMT 
detector response was not included, nor was the detector response to
the neutron component.  Within this approximation, however,
acceptable agreement between the two spectra over the energy
range covered by the detectors ($\approx 300$~\kev--15~\mev)
is found.  

\section{Operational lessons learned}

The beam test provided an opportunity to stress
the operation of the FEL energy recovery linac 
under high power, CW runs
of up to several hours in duration.  It was found that, while the machine tune
was somewhat critical, it was relatively 
straightforward to transmit the 
beam through the aperture with minimal loss.  

Stability of the ERL was generally good, but there were a number of 
of trips of the machine (See Fig.~\ref{fig:rad_data} for Run~4), mostly of short duration.
During CW operation,
a few outages up to of 10 to 30 minutes occurred 
when a cavity lost phase lock and had to be re-tuned, or when the
phases drifted and CW operation had to be interrupted to re-phase.
Phase drift was compounded 
by high outside ambient temperatures on some days, with correspondingly
higher vault ambient temperature, which affected 
stability of some components. Beam trips occurred for a variety of 
reasons, including 
an RF control module that needed replacement; instabilities
induced by ``pinging'' of a cryomodule with the beam, causing it to trip; and
phase-lock faults in the drive laser
associated with the high ambient temperatures.  
Operational stability
improved during the week as experience was gained.  

\section{Summary}

The DarkLight beam transmission test demonstrated the feasibility of
sustained CW operation of the JLab FEL ERL 
at high power (430 kW) with a 70~$\mu$m
beam spot on a 2~mm diameter target aperture similar to
that proposed for the DarkLight gas target.  Data on beam loss,
wall heating, and radiation backgrounds were obtained.
Average beam losses were less than 7~ppm for all aperture sizes.
Radiation backgrounds downstream of
the test fixture were measured and provide information
useful for design of the DarkLight detector.
Experience in operating the machine in this configuration
was obtained and techniques for machine tuning and beam diagnostics
developed.  The test demonstrated that 
the JLab FEL is capable of meeting
the needs of the DarkLight experiment. 
Beam optical qualities are already very good. 

Machine stability should be improved by planned upgrades to the FEL.
These upgrades include a new injector,
use of a smaller drive laser beam spot, 
reducing the bunch charge to 30~pC/pulse, 
and switching to a longer (80~cm) beta function.
Installation and operation of the DarkLight detector will allow
implementation of a transport system solution that---in addition to
providing an appropriately configured beam at the target---provides a
phase space exchange analogous to that normally used in the IR side of
the machine. This type of exchange has been demonstrated to suppress the
beam break-up (BBU) instability and to allow operation at very high current.

Measurements were obtained of photon and
neutron backgrounds produced by the FEL when running under 
conditions similar to what will be the case when a gas target is
installed for DarkLight.  Of several reference runs using the three
different apertures, the 7~hour, 2~mm aperture Run~4 was the most
stringent, delivering 121~C total.
On average during the latter half of Run~4, 
photon and neutron radiation backgrounds downstream of the test fixture
were
around 50~R/hr whole-body dose for photons and 
0.4 rem/hr whole-body dose for neutrons.
Numerous lessons were learned about machine operation at high current
and high power, including that decreases in radiation backgrounds 
were obtained by improved tuning of the machine during the course
of a run.  

A good deal was learned about ways to improve
performance and reduce backgrounds.
One example is beam break-up (BBU).  Usually BBU is eliminated
by use of a rotator, but during the transmission test a rotator normally used
for this purpose had been 
replaced by the test fixture and associated components.
An alternative was found: shifting
the beam energy to a slightly lower value eliminated the BBU.

%
%
%
%
\section{Acknowledgements}
We gratefully acknowledge the outstanding efforts of
both the staff of the Jefferson Laboratory to deliver the
high quality FEL beam and the staff of the MIT-Bates
Research and Engineering Center who designed, constructed
and delivered the test target assembly. The research
is supported by the United States Department of
Energy Office of Science. 

Notice: Authored by Jefferson Science Associates, LLC under U.S. DOE Contract No. DE-AC05-06OR23177.
The U.S. Government retains a non-exclusive, paid-up, irrevocable, world-wide license to publish or reproduce this manuscript for U.S. Government purposes. This work supported by the Commonwealth of Virginia and by DOE under contract DE-AC05-060R23177.

%
%
%
%
\section{Appendix: NaI/PMT photon measurements}
As part of the ambient radiation measurements,
additional spectra were taken under identical conditions except
for the amount of lead shielding: 2 inches and 4 inches surrounding
the NaI/PMT detectors. Analysis of these data provides information
that may be useful when planning any experiment to be
conducted in the vault's radiation environment.
\subsection{Effective attenuation}
A mono-energetic photon beam traversing material \footnote{For this discussion and unless otherwise specified, lead will be the attenuating material for photons.} is attenuated by several processes. At low energies, attenuation is dominated by the photoelectric effect. At higher energies the Compton effect is dominant. At still higher energies, pair production takes over. These effects overlap and the resulting attenuation can be expressed as:  
\begin{equation}
N_x \left( i \right)  =  N_v  \left( i \right) \exp \left[ - \left( \frac{\mu(i)}{\rho} \right) \rho x \right]
\end{equation}  
where $\mu(i)$ is the mass attenuation coefficient for photon energy $E(i)$, $\rho$, is the density of lead (11.35 gm/cm\textsuperscript{3}), and $x$ is the thickness of lead. A  photon spectrum can be considered to be the sum of mono-energetic photons of energy $E(i)$, where $i$ represents the channel number (or energy bin) over the detection range. Photons in each channel will be attenuated according to the above equation. 

\begin{figure}[htb]
\ifblackandwhite
\includegraphics[width=0.5\textwidth]{FIG9_Mass-attn-v2_black_and_white.pdf}\else
\includegraphics[width=0.5\textwidth]{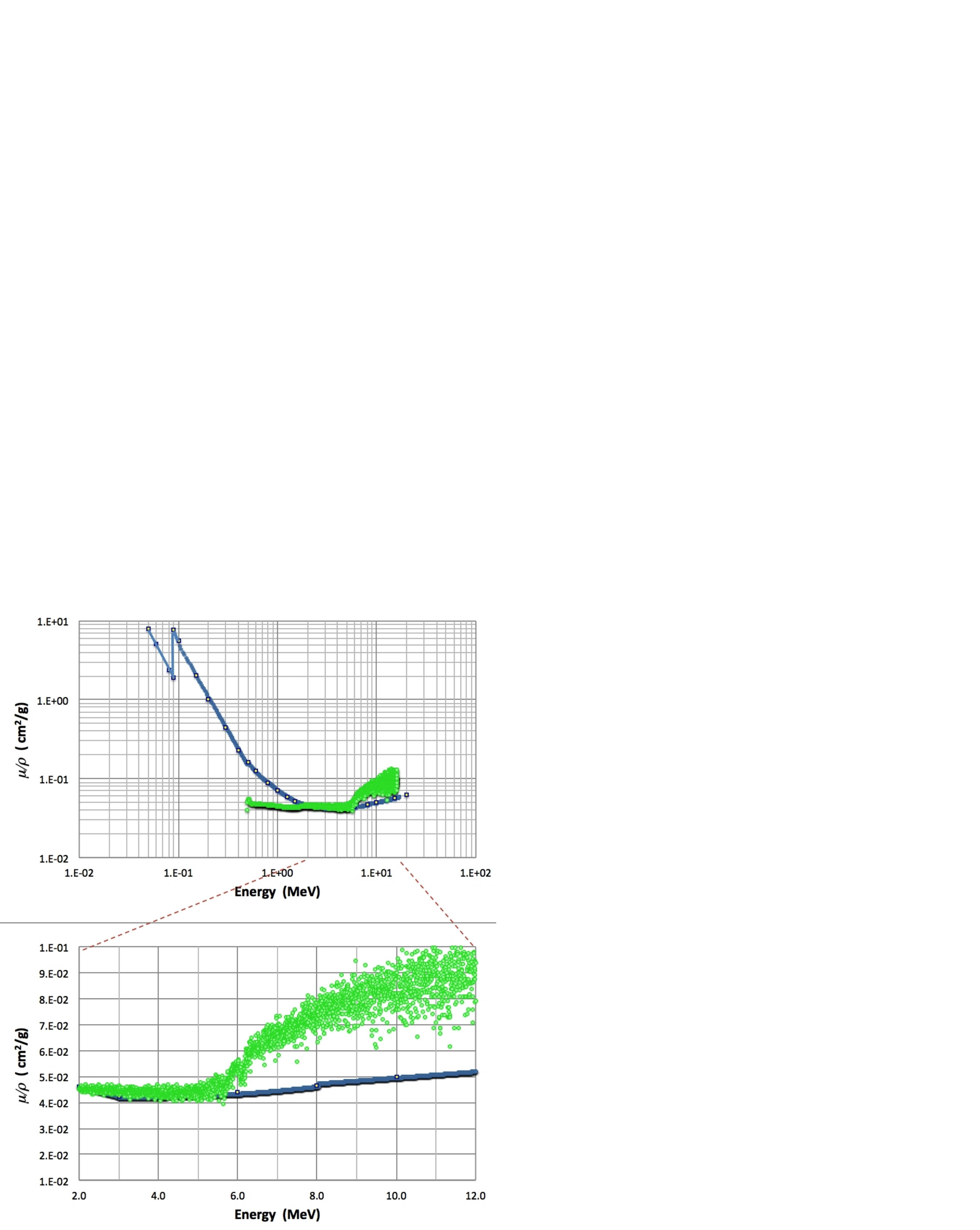}\fi
\caption{\label{Mass-attn}Mass attenuation coefficients for lead: (a) logarithmic scale; (b) linear scale. The curve with square points is from NIST tables~\protect\cite{Hubbell:2004ab}. The data (circles) are the effective mass attenuation coefficients derived from Run A and Run B. Note the significant increase in the value of the effective mass attenuation near 6 MeV photon energy. This suggests an additional mechanism for attenuation other that the PE-Comption-Pair production. The $(\gamma, ~n )$ threshold for lead is 6.2 MeV. }
\end{figure}

We measured the spectrum for two thicknesses $ x = 2$ and  4 inches: Run A and Run B, respectively. Note that the four-inch-thick box can be treated as two 2-inch thick boxes, one inside the other. The natural log of the ratio of the two spectra yields: 
\begin{equation}
\ln \left( \frac {N_4 \left( i \right)}{ N_2 \left( i \right) } \right)  = - \mu \left( i \right) \left( ~4
~{\rm inches} -  ~2 ~{\rm inches} \right)
\end{equation}  
or 
\begin{equation}
\left( \frac {\mu \left( i \right) }{\rho} \right) = \frac {1}{(2~{\rm inches})
 \rho~} \ln \left( \frac {N_4 \left( i \right)}{ N_2 \left( i \right) } \right) 
\end{equation}
Using the two spectra, Run A and Run B, in equation (3) yields the results shown in Fig.~\ref{Mass-attn}.

\subsection{Neutron spectrum derived from photon spectrum}

\begin{figure}[htb]
\ifblackandwhite
\includegraphics[width=0.5\textwidth]{FIG10_NeutronSpectrum_crop_black_and_white.pdf}\else
\includegraphics[width=0.5\textwidth]{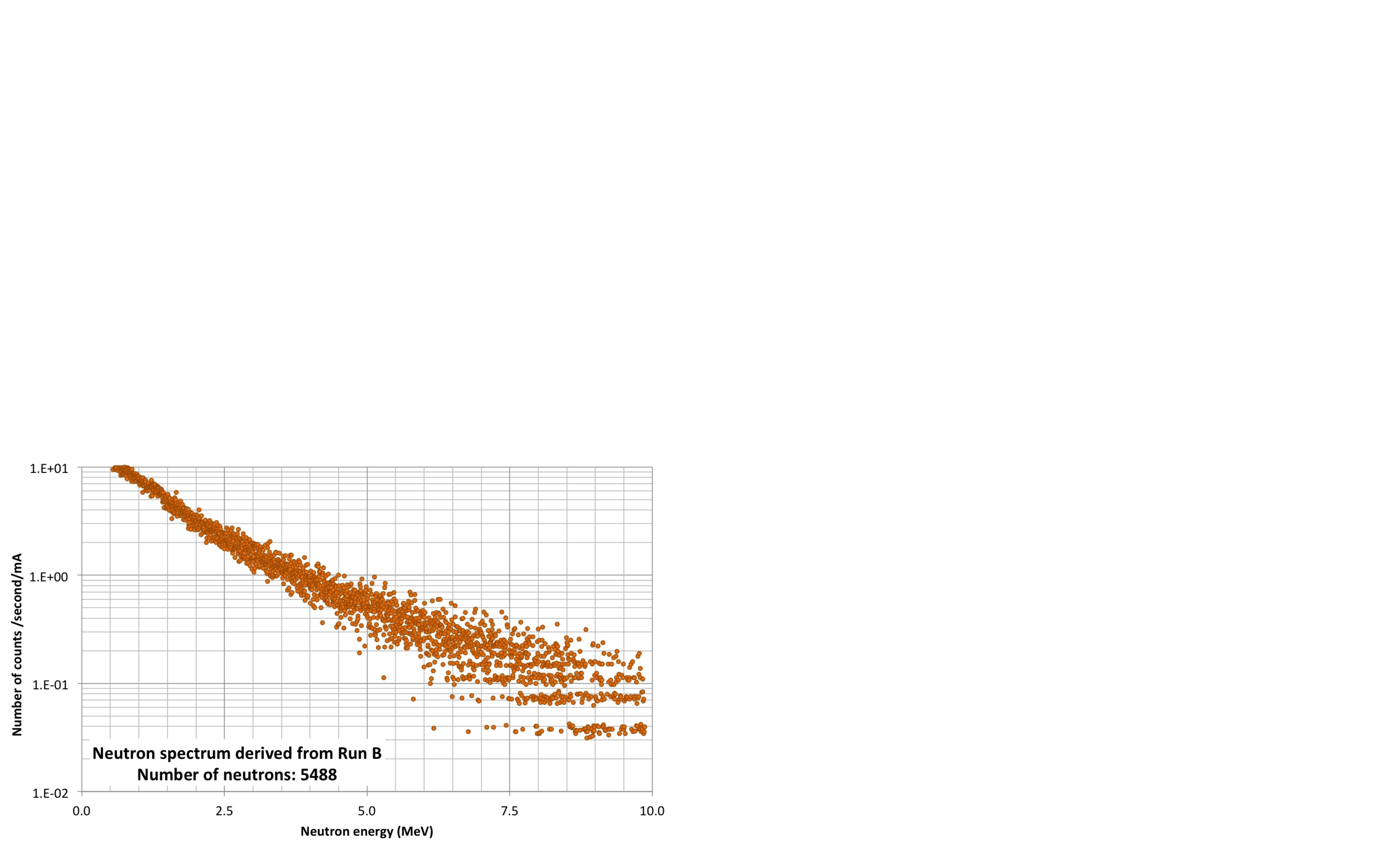}\fi
\caption{\label{NeutronSpectrum}Neutron spectrum derived from Run B. Each count represents a photon with energy greater than the $(\gamma, ~n)$ threshold for $^{208}$Pb. Thus there are at least 5488 gammas of energy greater than 6.0 MeV in the ambient vault background. }
\end{figure}

Up until the $(\gamma,~n)$ threshold, the effective attenuation of ambient photons is approximately constant. Assuming that this would be the case if there were no neutron production, then above threshold the liberated neutron would have kinetic energy equal to the difference between the photon energy and the binding energy of the neutron. Thus a neutron spectrum can be obtained from Run-B as shown in Fig.~\ref{NeutronSpectrum}. 
\subsection{511  keV Peak}
\begin{figure}[htb]
\ifblackandwhite
\includegraphics[width=0.5\textwidth]{FIG11_Run_4_e+e-_peak_crop_black_and_white.pdf}\else
\includegraphics[width=0.5\textwidth]{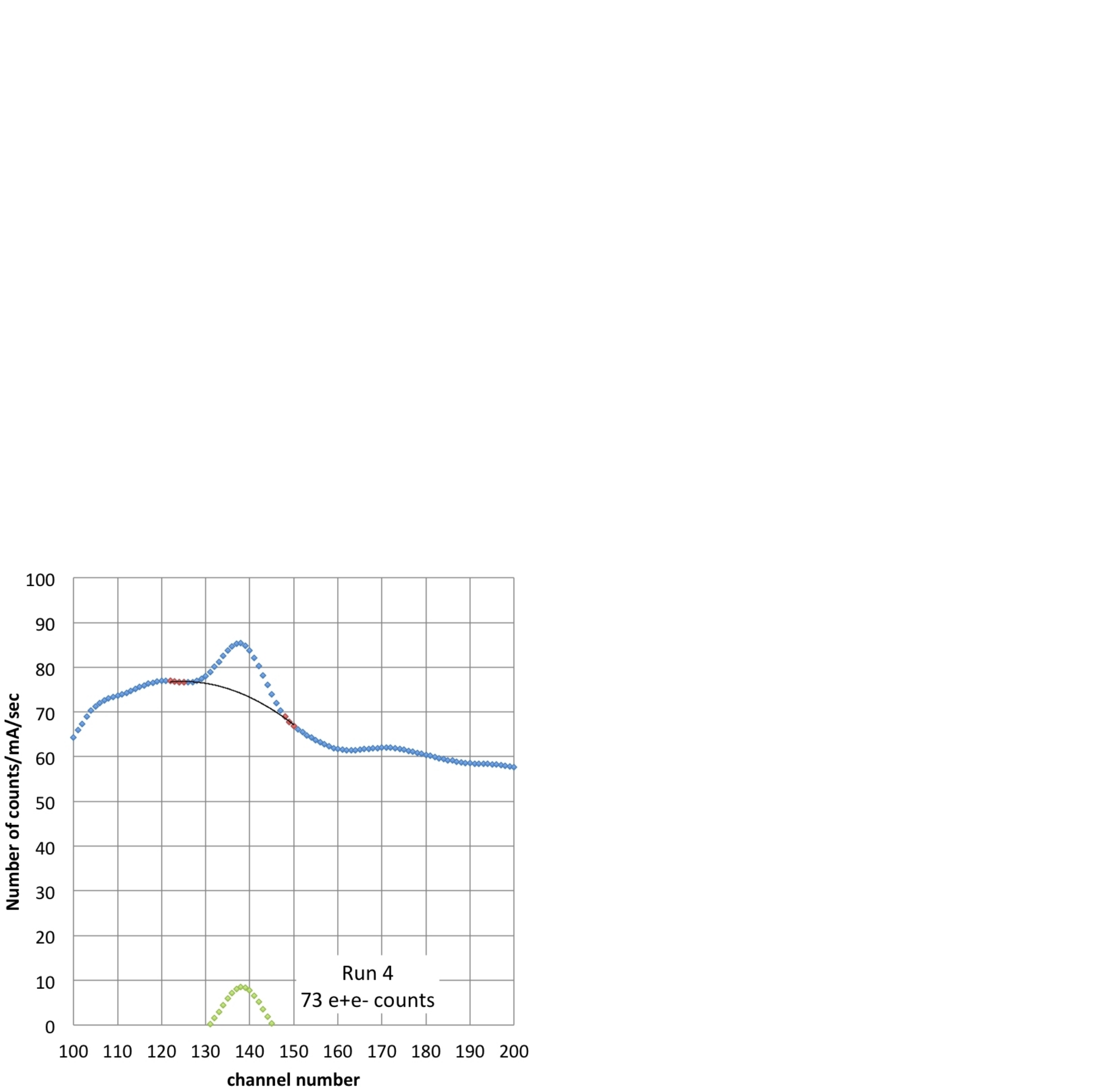}\fi
\caption{\label{RunM027-511keV}Run 4: the 511 keV annihilation peak. The number of counts in the peak near 511 keV is the sum of the counts above the quadratic background fit line (upper points). Since each 511 count (lower points) represents an electron-positron pair produced by a high energy gamma near the inner surface of the lead shielding, the number of counts in the peak (above the background line) is at least the number of high energy gammas that should be added to the sum of high energy gammas. That is, the vault radiation  above 6 MeV is  at least the sum of neutrons, the pair-producing gammas, and the photon spectrum.
}
\end{figure}

At the low energy end of the photon spectrum in Fig.~\ref{RunM027-511keV} there is a peak near 511 keV. This is due to pair production from high energy photons that have penetrated the shielding and pair produce at the inner surface of the lead box. Of course, pair production is occurring throughout the transit through the shielding but the lead shielding absorbs the 511 keV photons before they are seen in the NaI. 

\subsection{Final Observations}

The four-inch thick shielding can be thought of as a two-inch thick shielding box that is an analog Monte Carlo spectrum generator of an incident photon flux for the inner 2 inches of lead. Further, this takes into account the actual running conditions that are only approximated by a computer generated model.

In the FLUKA comparison with Run 4, the onset of $(\gamma, ~n)$ is seen in the change in slope of Run 4's data about 6 MeV. Note that the FLUKA spectrum does not show this effect.

Close comparision of the slopes before and after the peak of both spectra shows the effect of pile-up in Run 4's spectrum. The main reason for setting a lower level discriminator threshold so that only data above a few hundred keV, was to reduce the pile-up and dead time during collection of the spectra.

%

%
%
\bibliographystyle{model1-num-names}
\bibliography{bkgs2013}

\end{document}